\documentclass[12pt]{article}
\usepackage{amssymb,amsmath,graphicx}
\usepackage[cp1251]{inputenc}
\usepackage[english]{babel}

\graphicspath{{noiseimages/}{images/}}
\begin{document}

\begin{center}\large
  \textbf{Effective indirect multi-site spin-spin interactions \\
  in the s-d(f) model}
\end{center}

\begin{center}
  K.\,K.\,Komarov$^{a)}$, D.\,M.\,Dzebisashvili$^{a,b)}$
\end{center}

\begin{center}\small\it
  $^{a}$Kirensky Institute of Physics, Federal Research Center KSC SB RAS, 660036, Krasnoyarsk, Russian Federation\\
  $^{b}$Siberian State Aerospace University, Krasnoyarsk, 660014, Russian Federation
\end{center}

{\small Using the diagram technique for Matsubara Green's
functions it is shown that the dynamics of the localized spin
subsystem in the s-d(f) model can be described in terms of an
effective spin model with multi-site spin-spin interactions. An
exact representation of the action for the effective purely spin
model is derived as an infinite series in powers of s-d(f)
exchange interaction $J$. The indirect interactions of the 2-nd,
3-rd and 4-th order are discussed.}
\bigskip

\section*{1. Introduction}\label{sect:introduction}

Kondo lattice model or s-d(f) exchange model is widely used to
describe the correlation effects in metals (and their compounds)
with unfilled $d$- and $f$-shells.

The nature of the ground state in these systems is largely
determined by the result of competition between two interactions.
On the one hand, s-d(f) exchange coupling between spins of
itinerant s- and localized d(f)- electrons due to Kondo
fluctuations screens the localized spins and tends to form a
non-magnetic ground state \cite{Moshalkov1986}. In the opposite
direction acts indirect exchange (RKKY) interaction between spins
of d(f)-electrons \cite{Yosida1957}, trying to set a long-rang
magnetic order which is not necessarily a ferromagnetic or
antiferromagnetic. Correlation effects can lead to, for example,
helical magnetic structures \cite{Hamada1995}, with period that is
determined by the Fermi surface singularities
\cite{Yosida1962,Dzyaloshinskii1994}.

It is clear that the study of competition between different
effective interactions, that occur in the localized spin subsystem
and are caused by the same itinerant electrons, must be carried
out within the same unified approach.

The purpose of this paper is to derive such an effective
Hamiltonian (or rather action), which will allow to study the
localized spin subsystem in the s-d(f) model within the model with
only spin-spin interactions. At the same time all possible
spin-spin interactions that occur in different orders of
perturbation theory in $J$, should be determined by the same model
parameters: 1) the s-d(f)-exchange coupling constant $J$; 2) the
dispersion law of conduction electrons; 3) the filling degree of
the conduction band. Such an approach should allow to analyze all
types of exchange spin-spin interactions from a unified point of
view.

This problem is solved by integrating charge degrees of freedom
using the diagram technique for Matsubara Green's functions. It is
shown that in addition to the two-spin indirect exchange
interaction, which occurs in the second order in s-d(f)-parameter
$J$, in the following orders also appear terms describing, in
particular, the ring and biquadratic exchange interactions. The
essential point of all these interactions is the account for
retardation effects provided by the imaginary time dependence of
all the effective multi-site exchange parameters.
\bigskip

\section*{2. The Hamiltonian of the s-d(f) exchange model}\label{sect:hamilton}

The Hamiltonian of the Kondo lattice model (or s-d(f) exchange
model) can be written as a sum of two terms:
\begin{eqnarray}\label{eq:H}
  \hat H = \hat H_0 + \hat H_{int},
\end{eqnarray}
where
\begin{eqnarray}\label{eq:H_0&H_int}
  \hat H_0 = \sum_{k\alpha} (\varepsilon_{k} - \mu) c^+_{k\alpha} c_{k\alpha},~
  \hat H_{int} = \frac{J}{2} \sum_{f} c^+_f \tilde S_f c_f.
\end{eqnarray}
Operator $\hat H_0$ stands for the energy of noninteracting
current carriers (electrons or holes) with
dispersion~$\varepsilon_k$, $\mu$ -- chemical potential. Operator
${c^+_{k\alpha}}{(c_{k\alpha})}$ creates (annihilates) a particle
in the state with quasimomentum $k$ and spin projection
$\alpha=\pm1/2$.

The second term in (\ref{eq:H}) describes Kondo exchange
interaction between localized spins and itinerant quasiparticles.
The intensity of this interaction is defined by the constant $J$.
Operator ${\tilde S_f=\vec S_f\vec\sigma}$ is a product of a
localized spin vector operator $\vec S_f=(S^x_f,S^y_f,S^z_f)$ and
a vector ${\vec\sigma=(\sigma^x,\sigma^y,\sigma^z)}$ which is
formed of Pauli matrices. In definition (\ref{eq:H_0&H_int}) the
spinor notations:
\begin{eqnarray}\label{eq:C_c_S_S_Struktura_odnorod}
  c^+_f =
  \begin{pmatrix}
    c^+_{f\uparrow} & c^+_{f\downarrow}
  \end{pmatrix}\!,~
  c_f =
  \begin{pmatrix}
    c_{f\uparrow} \\ c_{f\downarrow}
  \end{pmatrix}\!,~
  \tilde S_f =
  \begin{pmatrix}
    S^z_f & S^-_f \\ S^+_f & -S^z_f
  \end{pmatrix}\!,~S^\pm_f=S^x_f\pm i S^y_f,
\end{eqnarray}
are used. The operators $c_f$ and $c_k$ are related to each other
by Fourier transformation: ${c_f=N^{-1/2}\sum_k e^{ikf} c_k}$.
\bigskip

\section*{3. Spin Green's functions and effective action}\label{sect:grin}

The thermodynamic properties of localized spin subsystem
conveniently studied on the basis of the diagram technique for
Matsubara Green's functions \cite{Matsubara1955}:
\begin{eqnarray}\label{eq:grin}
  G^{jj'}(f-f',\tau-\tau') = -\left\langle T_\tau ~\! \bar S^j_f(\tau)\bar S^{j'}_{f'}(\tau') \right\rangle, ~j, j'=\{x,y,z\}.
\end{eqnarray}
In this expression the spin operators are written down in the
Heisenberg representation: ${\bar S^j_f(\tau)= e^{\tau\hat H}
S^j_f e^{-\tau\hat H}}$, where $\hat H$ is the s-d(f) exchange
Hamiltonian (\ref{eq:H}), and $\tau$ -- imaginary time varying
within interval $(0, 1/T)$ ($T$ -- is the temperature). The
imaginary time ordering operator $T_\tau$ arranges all operators
on the right side of it in the descending order of the imaginary
time $\tau$ from left to right. Angle brackets in (\ref{eq:grin})
denote a thermodynamic average over the grand canonical ensemble
described by the Hamiltonian~$\hat H$.

As is known \cite{AGD1962}, in the interaction representation:
${S^j_f(\tau) = e^{\tau\hat H_0} S^j_fe^{-\tau\hat H_0}}$, the
expression (\ref{eq:grin}) transforms into:
\begin{eqnarray}\label{eq:G_ff}
  G^{jj'}(f-f',\tau-\tau') = -\left\langle T_\tau ~\! S^j_f(\tau) S^{j'}_{f'}(\tau') ~\! \mathfrak{S}(\beta) \right\rangle_{0}.
\end{eqnarray}
Here the scattering matrix $\mathfrak{S}(\beta)$ is defined by the
formula:
\begin{eqnarray}\label{eq:Smatrix}
  \mathfrak{S}(\beta) = T_\tau\exp\left\{-\int_0^\beta \mathrm{d}\tau\hat H_{int}(\tau) \right\},
\end{eqnarray}
where in the exponent under the integral stands the interaction
operator (\ref{eq:H_0&H_int}) in the interaction representation.
Lower index "0", on the right of the angle brackets in
(\ref{eq:G_ff}) indicates that the thermodynamic averaging is over
the ensemble of non-interacting spin and fermion systems. Besides,
when expanding the thermodynamic averages using Wick's theorem
only connected diagrams should be considered.

Note that for calculating the Green's function (\ref{eq:G_ff})
along with the usual diagram technique \cite{AGD1962} it is
necessary to use the diagram technique for spin operators
\cite{Izyumov1974book} as well. The expanding of a
$T_\tau$-ordered average of a product of spin $S$- and fermion
$c$-operators can be divided into two stages: first, only
$c$-operators are paired, and then the Wick's theorem is applied
to the remaining spin operators. Formally this can be written down
as follows:
\begin{eqnarray}\label{eq:SScc}
  \left\langle T_\tau ~\! S_1 \dots S_l ~\! c_1\dots c^+_m \right\rangle_{0} =
   \Big\langle T_\tau ~\! S_1 \dots S_l ~\! \left\langle T_\tau ~\! c_1 \dots c^+_m \right\rangle_{c0}~\Big\rangle_{S0}.
\end{eqnarray}
Internal thermodynamic average on the right side of the equation
(\ref{eq:SScc}) with index "$c0$" denotes averaging only over an
ensemble of non-interacting fermions. External $T_\tau$-ordered
averaging, indicated by the index "$S0$", should be done using the
ensemble of non-interacting spin subsystems. Applying equation
(\ref{eq:SScc}) to the definition of the Green's function
(\ref{eq:G_ff}) we can write:
\begin{eqnarray}\label{eq:G_ff2}
  G^{jj'}(f-f',\tau-\tau') = -\left\langle T_\tau ~\! S^j_f(\tau) S^{j'}_{f'}(\tau') ~\! \mathfrak{S}_S(\beta) \right\rangle_{S0},
\end{eqnarray}
where the effective scattering matrix $\mathfrak{S}_S(\beta)$ is defined by the expression:
\begin{eqnarray}\label{eq:Seff}
  \mathfrak{S}_S(\beta) = \left\langle \mathfrak{S}(\beta) \right\rangle_{c0}.
\end{eqnarray}

To obtain the effective purely spin model first we should pair all
the $c$-operators according to the Wick's theorem in each order of
the $\mathfrak{S}$-matrix expansion in powers of the coupling
constants $J$. After that the result must be rewritten as a
$T_\tau$-ordered exponent. The argument of the exponent will
determine the required effective interaction in the subsystem of
localized spins. Note that the described here scheme of
integrating over the charge degrees of freedom in the s-d(f) model
in some sense is similar to the proof of equivalence between
diagrammatic expansion for the Green's function of localized
$f$-electrons in the periodic Anderson model and diagrammatic
expansion of the fermion Green's function in the Hubbard model
\cite{Moskalenko1997,VDB2006}.

\begin{figure}[ht]
  \begin{center}
    \includegraphics[width=400pt, height=300pt, angle=0, keepaspectratio]{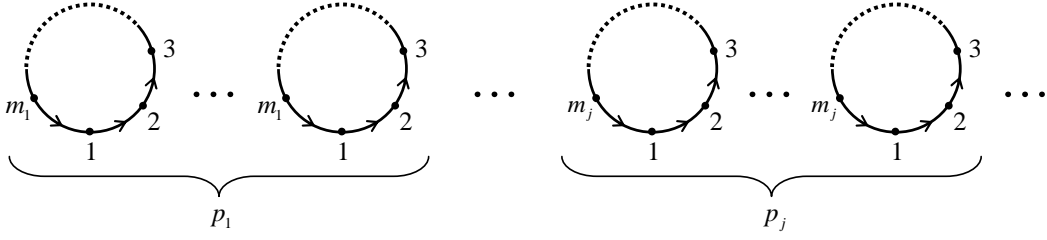}
    \caption{\label{fig:loops}
The $n$-th order diagrams, arising due to pairing of all
$c$-operators in the expansion of $\langle
\mathfrak{S}(\beta)\rangle_{c0}$, can be represented as a set of
fermionic loops. The order of each loop $m_j$ is determined by the
number of vertices represented by points and the same number of
lines with arrows denoting the fermion propagator. Each vertex
corresponds to the spin operator. If the number of vertices of the
same order $m_j$ is denoted by $p_j$, then:
$p_1m_1+p_2m_2+\dots=n$.}
  \end{center}
\end{figure}

Let us consider for the average $\langle
\mathfrak{S}(\beta)\rangle_{c0}$ the diagrams of $n$-th order with
respect to the constant $J$. Since each term in $\hat H_{int}$
consists of one spin operator and two second-quantization
operators ($c^+$ and $c$), the structure of diagrams emerging
after all possible pairings of $c$-operators is characterized by a
some number of fermionic loops (see figure \ref{fig:loops}). The
order of each loop is determined by the number of fermionic lines,
corresponding to the fermion Green's functions
${G^{(0)}_k(\tau-\tau') = -\langle T_\tau~c_k(\tau) c^+_k(\tau')
\rangle_0}$, and the same number of vertices. In contrast to the
conventional diagram technique in this case each vertex is related
not to a function (or a number) but to the spin operator.
Therefore the $n$-th order diagram corresponds to a product of $n$
spin operators entering to the $T_\tau$-ordered thermodynamic
average $\langle T_\tau\dots\rangle_{S0}$.

Let us denote by $p_j$ the number of loops of the same order
$m_j$, where the index $j$ enumerates all loops of different
order. Then obviously $n=p_1m_1+p_2m_2+\dots$. Until the spin
operators are not paired the analytical expression corresponding
to each $n$-th order diagram can be represented as a product of
some multipliers. Each multiplier is related to a single fermionic
loop. Note that in the case of further application of the spin
diagram technique \cite{Izyumov1974book} there should be taken
into account only such pairings of spin operators when all
fermionic loops are connected graphically with elements of the
spin diagram technique.

The fermion loop of the $m_j$-th order is formed due to $m_j$
operators $\hat H_{int}$. Let's use the index "$L$" in angle
brackets $\langle \dots\rangle_L$ to indicate the fact that when
pairing all the $c$-operators in the $T_\tau$-ordered average of
$m_j$ operators $\hat H_{int}$ there should be taken into account
only one-loop diagrams of $m_j$-th order. Then the analytical
contribution to the one loop of $m_j$-th order will be determined
by the average: $\langle T_\tau \hat H_{int}(\tau_1)\dots \hat
H_{int}(\tau_{m_j}) \rangle_{Lc0}$. Strictly speaking this average
gives rise to $(m_j-1)!$ equivalent fermionic loops of $m_j$-th
order. Using the definition of the average $\langle T_\tau
\dots\rangle_{Lc0}$ the expansion of $\langle \mathfrak{S}(\beta)
\rangle_{c0}$ can be written as:
\begin{eqnarray}\label{eq:Sser}
  \bigl\langle \mathfrak{S}(\beta) \bigl\rangle_{c0} =
    T_\tau\sum_n\frac{(-1)^n}{n!} \sum_{{p_1, p_2\dots\atop{m_1, m_2\dots}}\atop{(p_1m_1+\dots=n)}} A(p_1m_1; p_2m_2; \dots) \times
    ~~~~~~~~~\nonumber\\
    \times\prod_j\int \mathrm{d}\tau^{(j)}_{1}\dots \mathrm{d}\tau^{(j)}_{m_1}
    \bigl\langle T_\tau\hat H_{int}(\tau^{(j)}_{1})\dots\hat H_{int}(\tau^{(j)}_{m_1})\bigl\rangle_{Lc0}.
\end{eqnarray}
In derivation the formula (\ref{eq:Sser}) in each $n$-th order of
$\hat H_{int}$ the thermodynamic average $\langle T_\tau
\dots\rangle_{c0}$ was represented as a sum of products of
averages $\langle T_\tau \dots\rangle_{Lc0}$. The second sum in
(\ref{eq:Sser}) takes into account all possible partitions of $n$
operators $\hat H_{int}$ into $p_1+p_2+\dots$ groups. Each group
corresponds to a single fermionic loop and the equality:
$p_1m_1+p_2m_2+\dots=n$ must be satisfied. The index $j$ in the
product $\prod_j$ runs all groups (loops) at a fixed partition.
Because the averaging in the $\langle
\mathfrak{S}(\beta)\rangle_{c0}$ is carried out only over the
charge degrees of freedom, the remaining spin operators must be
still ordered by the imaginary time. This justifies the presence
of operator $T_\tau$ in the right side of (\ref{eq:Sser}).
Combinatorial factor $A(p_1m_1; p_2m_2; \dots)$ takes into account
the equivalent contributions arising at each possible partitioning
of $n$ operators $\hat H_{int}$ in groups.

It can be shown (cf. \cite{AGD1962}) that:
\begin{eqnarray}\label{eq:An}
  A(p_1m_1; p_2m_2; \dots) = \frac{n!}{p_1!~\!(m_1!)^{p_1}~\!p_2!~\!(m_2!)^{p_2}\dots}.
\end{eqnarray}

Introducing notations:
\begin{eqnarray}\label{eq:Xi_m_int}
  \Xi_m=\frac{(-1)^{m+1}}{m!}\int_0^\beta \mathrm{d}\tau_{1}\dots \mathrm{d}\tau_{m}
  \left\langle T_\tau\hat H_{int}(\tau_{1})\dots \hat H_{int}(\tau_{m})\right\rangle_{Lc0},
\end{eqnarray}
the expression (\ref{eq:Sser}) can be rewritten as:
\begin{eqnarray}\label{eq:Sser2}
  \bigl\langle \mathfrak{S}(\beta)\bigl\rangle_{c0} \equiv
  \mathfrak{S}_S(\beta) = T_\tau \sum_{p_1,p_2\dots} \frac{1}{p_1!}(-\Xi_1)^{p_1} \frac{1}{p_2!}(-\Xi_2)^{p_2} \dots =
  T_\tau\exp\left\{-\Xi_1-\Xi_2-\dots \right\}.
\end{eqnarray}

Using the explicit form of $\hat H_{int}$ carry out in
(\ref{eq:Xi_m_int}) pairings of all the $c$-operators. Then
introducing the variable $x_j=(\vec R_{f_j},\tau_j)$ we obtain
expression for the operator $\Xi_m$ in the Wannier representation:
\begin{eqnarray}\label{eq:Xi_m}
  \Xi_m = \frac{1}{m} \left(\frac{J}{2}\right)^{m} \int \mathrm{d}x_1\dots \mathrm{d}x_m
         ~G^{(0)}(x_1-x_2) \dots G^{(0)}(x_m-x_1)~{\rm Sp} \left\{\tilde S(x_1)\dots \tilde S(x_m)\right\},
\end{eqnarray}
where the integral over $\mathrm{d}x$ denotes the operation
$\int_0^\beta \mathrm{d}\tau\sum_f$ and the trace is calculated
over the Pauli matrix indices.

\begin{figure}
  \begin{center}
    \includegraphics[width=290pt, keepaspectratio]{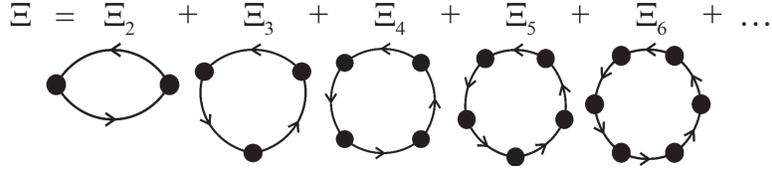}
    \caption{\label{fig:Xi_sum}
The effective action $\Xi$ can be expressed as an infinite series
of terms $\Xi_n$. Each term $\Xi_n$ in the diagrammatic
representation corresponds to a single loop of $n$-th order in the
exchange coupling constant $J$ and describes the effective
$n$-site spin-spin interaction. The lines with arrows represent
bare fermion Green's functions and each vertex, shown with a bold
circle, corresponds to a spin operator $\tilde S$.}
  \end{center}
\end{figure}

From the formula (\ref{eq:Sser2}) it follows that the effective
interactions in the localized spin subsystem are determined by the
structure of the series:
\begin{eqnarray}\label{eq:Xi}
  \Xi = \sum^{\infty}_{m=1}\Xi_m.
\end{eqnarray}
The effective action $\Xi$ describes all possible multi-site
spin-spin interactions in the localized spin subsystem in the
arbitrarily order of the coupling constant $J$. The partial action
$\Xi_m$ determines the $m$-th order interactions and
diagrammatically can be represented as a loop with $m$ lines,
corresponding to propagators $G^{(0)}$, and $m$ vertices related
to spin operators $\tilde S$ (see figure \ref{fig:Xi_sum}). It can
be seen that all effective interactions in $\Xi$ take into account
retardation effects. This is due to imaginary time dependence of
all the effective interactions in each term (\ref{eq:Xi_m}) of the
series for $\Xi$.
\bigskip

\section*{4. Effective interactions of the 2nd, 3rd and 4th order in $J$}\label{sect:eff_int}

Let us consider the first several terms of the series for $\Xi$.
The first term with $n=1$ is zero, because ${\mathrm{Sp}\{\tilde
S(x_1)\} = \sum_j S^j(x_1) \mathrm{Sp}\{\sigma^j\}}$, and
${\mathrm{Sp}\{\sigma^j\}=0}$ at any $j=x,y,z$.

\begin{figure}[h]
  \begin{center}
    \includegraphics[width=100pt, keepaspectratio]{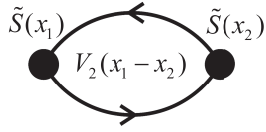}
    \caption{\label{fig:Xi_2}
The action $\Xi_2$ can be represented by a loop of the second
order in the exchange coupling constant $J$.}
  \end{center}
\end{figure}

To calculate the operator $\Xi_2$ one should to derive the trace
${\mathrm{Sp}\left\{\tilde S(x_1) \tilde S(x_2)\right\}}$. Using
the identity:
${\sigma^i\sigma^j=\delta_{ij}+i\varepsilon_{ijl}\sigma^l}$, where
$\varepsilon_{ijl}$ -- Levi-Civita antisymmetric tensor we find:
\begin{eqnarray}
  \mathrm{Sp}\left\{\tilde S(x_1) \tilde S(x_2)\right\} =
     2\left(\vec S(x_1)\vec S(x_2)\right).
\end{eqnarray}
Then the operator $\Xi_2$ takes the form:
\begin{eqnarray}\label{eq:Xi_2}
  \Xi_2 = \int \mathrm{d}x_1 \mathrm{d}x_2~V_2(x_1-x_2)~\vec S(x_1) \vec S(x_2),
\end{eqnarray}
where the effective interaction between spins is defined via a
polarization loop (see figure \ref{fig:Xi_2}):
\begin{eqnarray}\label{eq:V_2}
  V_2(x_1-x_2)=\left( \frac{J}{2} \right)^{2} G^{(0)}(x_1-x_2) G^{(0)}(x_2-x_1).
\end{eqnarray}

From equation (\ref{eq:Xi_2}) it follows, that at $f_1 \ne f_2$
the second order effective action $\Xi_2$ describes indirect
exchange interaction of two localized spins through the subsystem
of itinerant electrons. However, in contrast to the usual
RKKY-interaction here the retardation effects, caused by
$\tau$-dependence of $V_2$, are taken into account. Note that in
the $\Xi_2$ there is also a term with $f_1=f_2$. The Fourier
transform of the indirect exchange interaction (\ref{eq:V_2}) has
the form:
\begin{eqnarray}\label{eq:V_2_k}
  V_2(k, i\omega_m) = \left( \frac{J}{2} \right)^{2} \chi_0(k,i\omega_m),
\end{eqnarray}
where
\begin{eqnarray}\label{eq:chi}
  \chi_0(k, i\omega_m) = \frac1N \sum_q \frac{f_q-f_{q+k}}{i\omega_m+\varepsilon_q-\varepsilon_{q+k}}
\end{eqnarray}
is the Lindhard susceptibility in the Matsubara representation,
${\omega_m=2m\pi T}$ with ${m\in\mathbb{Z}}$, and
${f_q=\left(\exp\left\{(\varepsilon_q-\mu)/T\right\}+1\right)^{-1}}$
is the Fermi-Dirac distribution function. The intensity of the
exchange interaction is largely determined by the properties of
the itinerant subsystem. The well known RKKY-interaction follows
from (\ref{eq:chi}) at ${\omega_m\equiv 0}$.

\begin{figure}
  \begin{center}
    \includegraphics[width=85pt, keepaspectratio]{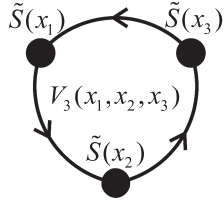}
    \caption{\label{fig:Xi_3}
Diagrammatic representation of the third order effective action
$\Xi_3$.}
  \end{center}
\end{figure}

Calculating the 3rd order effective action $\Xi_3$ we obtain (see
also figure \ref{fig:Xi_3}):
\begin{eqnarray}\label{eq:Xi_3}
  \Xi_3 = \int \mathrm{d}x_1 \mathrm{d}x_2 \mathrm{d}x_3~V_3(x_1, x_2, x_3)~\vec S(x_1) \cdot \left(\vec S(x_2) \times \vec S(x_3)\right),
\end{eqnarray}
where
\begin{eqnarray}\label{eq:V_3}
  V_3(x_1,x_2,x_3)= i \frac23 \left( \frac{J}{2} \right)^{3} G^{(0)}(x_1-x_2) G^{(0)}(x_2-x_3) G^{(0)}(x_3-x_1).
\end{eqnarray}
The formula (\ref{eq:Xi_3}) describes the three-spin interactions
in the form of a mixed product of three spin operators. This type
of interaction was first considered in \cite{Brown1984} but
without taking into account the retardation effects.

It is evident that the interaction (\ref{eq:Xi_3}) favors the
chiral order in the magnetic subsystem. In this regard we note
that in the paper \cite{Tatara2002} the third order corrections to
the Hall conductivity due to s-d(f) exchange interaction were
shown to give rise the anomalous Hall effect provided that non
trivial spin configuration (chirality) is formed in the spin
subsystem. Interestingly the structure of the expression for the
Hall conductivity obtained in \cite{Tatara2002} is similar to that
of (\ref{eq:Xi_3}).

\begin{figure}
  \begin{center}
    \includegraphics[width=110pt, keepaspectratio]{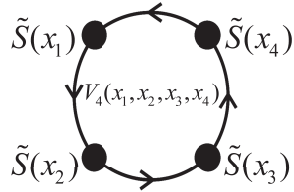}
    \caption{\label{fig:Xi_4}
      Fourth order effective action $\Xi_4$ in the diagrammatic representation.}
  \end{center}
\end{figure}

The operator $\Xi_4$, which determines the effective spin-spin
interactions in the fourth order in the coupling constant $J$,
after calculating the trace ${\mathrm{Sp}\{\tilde S(x_1) \tilde
S(x_2) \tilde S(x_3) \tilde S(x_4)\}}$ takes the form (see also
figure \ref{fig:Xi_4}):
\begin{eqnarray}\label{eq:Xi_4}
  \Xi_4 = \int \mathrm{d}x_1 \mathrm{d}x_2 \mathrm{d}x_3 \mathrm{d}x_4~V_4(x_1,x_2,x_3,x_4)~\left(\vec S(x_1) \vec S(x_2)\right)\cdot\left(\vec S(x_3) \vec S(x_4)\right),
\end{eqnarray}
where
\begin{eqnarray}\label{eq:V_4}
  V_4(x_1, x_2, x_3, x_4) = \frac12 \left( \frac{J}{2} \right)^{4}
                            \left[\right. G^{(0)}(x_1-x_2) G^{(0)}(x_2-x_3) G^{(0)}(x_3-x_4) G^{(0)}(x_4-x_1)-\nonumber\\
                                        -~G^{(0)}(x_1-x_3) G^{(0)}(x_3-x_2) G^{(0)}(x_2-x_4) G^{(0)}(x_4-x_1)+\nonumber\\
                                        +~G^{(0)}(x_1-x_4) G^{(0)}(x_4-x_3) G^{(0)}(x_3-x_2) G^{(0)}(x_2-x_1)\left.\right].
\end{eqnarray}

The terms of the expression (\ref{eq:Xi_4}) with unequal indices
of sites $f_j$ ($j=1,\dots,4$) correspond to the four-spin
exchange interactions. Among them there are, in particular,
interactions describing ring exchange of four spins located, for
example, at the square plaquette vertices. For the first time the
ring exchange (without retardation effects) was obtained from the
Hubbard model at half filling in the fourth order of the
perturbation theory in the parameter $t/U$, where $t$ is the
tunneling integral, and $U$ is the Coulomb repulsion energy of two
electrons at the same site \cite{Takahashi1977}. The four-spin
ring exchange interaction was involved to explain the magnetic
ordering features in the quantum crystal $^3$He \cite{Roger1983}.
In the paper \cite{Roger1989} it was argued that ring exchange is
important to describe magnetic properties of cuprate
high-temperature superconductors. Effect of ring exchange
interaction on the superconductivity in cuprates was investigated
in \cite{Shneider2012}.

At pairwise coincident site indices: $f_3=f_1$ and $f_4=f_2$, in
the sum (\ref{eq:Xi_4}) there are terms that are responsible for
biquadratic exchange interaction. These interactions were first
used in \cite{Harris1963} for explaining the paramagnetic
resonance on Mn ions in the compound MnO. Besides, the biquadratic
exchange interaction is essential in multilayer magnetic systems
\cite{Slonczewski1993}.

\section*{5. Conclusion}\label{sect:result}

The paper presents a method of deriving all possible kinds of
effective indirect interactions between the localized spins due to
s-d(f) exchange coupling of these spins with the subsystem of
itinerant electrons. After integrating over charge degrees of
freedom in the s-d(f) exchange model an exact representation for
an action of a purely spin model is obtained. Using this action
allows to study the spin subsystem in the s-d(f) model in the
framework of effective purely spin model. The important point of
this model is that all effective interactions take into account
retardation effects. Although explicit expressions are written
only for \mbox{two-,} three- and four-spin interactions, the
formula (\ref{eq:Xi_m}) allows to generate multi-site spin-spin
interactions in the arbitrary order in the s-d(f) exchange
coupling constant $J$.
\bigskip

The work was supported by the Russian Foundation for Basic
Research (project nos. 16-42-240435 and 15-42-04372).


\begin{thebibliography}{100}
  \bibitem{Moshalkov1986}\label{bib:Moshalkov1986}
    {V.V. Moschalkov, N.B. Brandt}, {UFN} {\bf 149} (1986) 585.
  \bibitem{Yosida1957}\label{bib:Yosida1957}
    {M.A. Ruderman, C. Kittel}, {Phys. Rev.} {\bf 96} (1954) 99;
    {K. Yosida}, {Phys. Rev.} {\bf 106} (1957) 893;
    {T. Kasuya}, {Prog. Theor. Phys.} {\bf 16} (1956) 45.
  \bibitem{Hamada1995}\label{bib:Hamada1995}
    {M. Hamada, H. Shimahara}, {Phys. Rev. B.} {\bf 51} (1995) 3027.
  \bibitem{Costa2015}\label{bib:Costa2015}
    {N. de C. Costa, J.P. de Lima, R.R. dos Santos}, {arXiv: 1506.00890v1 [cond-mat.str-el]} (2015).
  \bibitem{Yosida1962}\label{bib:Yosida1962}
    {K. Yosida, A. Watabe}, {Prog. Theor. Phys.} {\bf 28} (1961) 361.
  \bibitem{Dzyaloshinskii1994}\label{bib:Dzyaloshinskii1994}
    {I.E. Dzyaloshinskii}, {ZhETF} {\bf 47} (1964) 336.
  \bibitem{Matsubara1955}\label{bib:Matsubara1955}
    {T. Matsubara}, {Prog. Theor. Phys.} {\bf 14} (1955) 351.
  \bibitem{AGD1962}\label{bib:AGD1962}
    {A.A. Abrikosov, L.P. Gorkov, I.E. Dzyaloshinskiy}, {Methods of quantum
    field theory in statistical physics. -- Moscow: Fizmatgiz, 1962. -- 444 p.}
  \bibitem{Izyumov1974book}\label{bib:Izyumov1974book}
    {Yu.A. Izyumov, F.A. Kassan-ogly, Yu.N. Skryabin},
    {Field methods in the theory of ferromagnetism. -- Moscow: Nauka, 1974. S. -- 224}
  \bibitem{Moskalenko1997}\label{bib:Moskalenko1997}
    {V.A. Moskalenko}, {TMF} {\bf 110} (1997) 308.
  \bibitem{VDB2006}\label{bib:VDB2006}
    {V.V. Val'kov, D.M. Dzebisashvili}, {Pis'ma v ZhETF} {\bf 84} (2006) 251.
  \bibitem{Brown1984}\label{bib:Brown1984}
    {H.A. Brown}, {JMMM} {\bf 43} (1984) L1-L2.
  \bibitem{Tatara2002}\label{bib:Tatara2002}
    {G. Tatara, H. Kawamura}, {JPSJ} {\bf 71} (2002) 2613.
  \bibitem{Takahashi1977}\label{bib:Takahashi1977}
    {M. Takahashi}, {J. Phys. C: Solid State Phys.} {\bf 10} (1977) 1289.
  \bibitem{Roger1983}\label{bib:Roger1983}
    {M. Roger, J.H. Hetherington, J.M. Delrieu}, {Rev. Mod. Phys.} {\bf 55} (1983) 1-64.
  \bibitem{Roger1989} \label{bib:Roger1989}
    {M. Roger, J.M. Delrieu}, {Phys. Rev. B} {\bf 39} (1989) 2299.
  \bibitem{Shneider2012}\label{bib:Shneider2012}
    {E.I. Shneyder, S.G. Ovchinnikov, A.V. Shnurenko}, {Pis'ma v ZhETF} {\bf 95} (2012) 211.
  \bibitem{Harris1963}\label{bib:Harris1963}
    {E.A.Harris, J.Owen}, {Phys. Rev. Lett.} {\bf 11} (1963) 9.
  \bibitem{Slonczewski1993} \label{bib:Slonczewski1993}
    {J.C. Slonczewski}, {J. Appl. Phys.} {\bf 73} (1993) 5957.
\end{thebibliography}
\end{document}